# From academic to media capital: To what extent does the scientific reputation of universities translate into Wikipedia attention?

Wenceslao Arroyo-Machado[1], Adrián A. Díaz-Faes[2]*, Enrique Herrera-Viedma[1], and Rodrigo Costas[3,4]

wences@ugr.es; diazfaes@csic.es; viedma@ug.es; rcostas@cwts.leidenuniv.nl

[1] Department of Information and Communication Sciences, University of Granada

[2] INGENIO (CSIC-UPV), Universitat Politècnica de València, Camino de Vera s/n, 46022 Valencia, Spain

[3] Centre for Science and Technology Studies (CWTS), Leiden University

[4] DSI-NRF SciSTIP, Stellenbosch University

* Corresponding author

**Abstract**

Universities face increasing demands to improve their visibility, public outreach, and online presence. There is a broad consensus that scientific reputation significantly increases the attention universities receive. However, in most cases estimates of scientific reputation are based on composite or weighted indicators and absolute positions in university rankings. In this study, we adopt a more granular approach to assessment of universities' scientific performance using a multidimensional set of indicators from the Leiden Ranking and testing their individual effects on university Wikipedia page views. We distinguish between international and local attention and find a positive association between research performance and Wikipedia attention which holds for regions and linguistic areas. Additional analysis shows that productivity, scientific impact, and international collaboration have a curvilinear effect on universities' Wikipedia attention. This finding suggests that there may be other factors than scientific reputation driving the general public's interest in universities. Our study adds to a growing stream of work which views altmetrics as tools to deepen science-society interactions rather than direct measures of impact and recognition of scientific outputs.

**From academic to media capital: To what extent does the scientific reputation of universities translate into Wikipedia attention?**

### Introduction

Demonstrating societal impact and generating scientific results that respond to societal needs are of increasing concern to governments and public funding bodies (Penfield et al., 2014). The emergence of social media and the so-called altmetrics—web-based metrics on how people interact with research outputs (Priem et al., 2010)—have placed great expectations on developing measures that account for the impact of scientific work beyond academic boundaries (Sugimoto et al., 2017; Torres-Salinas et al., 2013). Until recently, altmetrics research focused mostly on application of citations models to social media (Haustein et al., 2016; Robinson-Garcia et al., 2018) as if the measurement of impact using citations and altmetrics were somehow equivalent (Ravenscroft et al., 2017). However, the evidence provided by this literature suggests that altmetrics as raw metrics are poor proxies for scientific or societal impact (Bornmann et al., 2019; Kassab et al., 2020), and that with some exceptions (Thelwall, 2018), they are not predictors of subsequent citations (Costas et al., 2015; Haustein et al., 2015). These findings have triggered a new wave of research that view altmetrics from a science-society lens rather than as direct measures of impact and recognition (Costas et al., 2020; Díaz-Faes et al., 2019; Wouters et al., 2019). This shift is enabling novel methodological approaches to capture and characterize social media activity from more interactive perspectives (Arroyo-Machado et al., 2021; Fang et al., 2022). In these approaches, the focus extends beyond research objects—i.e. papers and research products—to include news media and other means —e.g. blogs, websites—that allow societal actors to share and discuss research, its epistemic features—e.g. research topics, disciplines, schools of thought, etc.—and types of engagement with research objects—e.g. access, appraisal, and application (Alperin et al., 2023; Costas et al., 2020; Haustein et al., 2016).

In this study, we contribute to this new direction by providing a more granular understanding of the extent to which universities' scientific reputation is associated with the attention they attract on social media. We regard university reputation as a form of symbolic capital based on accumulated recognition, prestige, and influence (Bourdieu, 2004). University reputation depends on science-related aspects (e.g. scientific authority of its members) and other forms of economic, social, and cultural capital including financial resources, relations and networks, and technical expertise (Archer et al., 2015). We measure university scientific reputation using a multidimensional set of performance indicators from the Leiden Ranking and level of social media attention based on Wikipedia page views (i.e. traffic data).

Social media has become mainstream organizational communication and universities are aware of this shift (Ann Voss & Kumar 2013). Combined with the increasingly marketized environment in which universities operate, this is requiring universities to increase their visibility and public outreach (McIntosh et al., 2022). Although scientific reputation plays a crucial role in the online attention received by a university, this relationship is not straightforward since research capacity and scientific impact do not translate automatically into greater online visibility (Holmberg, 2015; Rybiński & Wodecki, 2022). This issue constitutes the empirical background to the present research.

<div align="center">**Background**</div>

**Wikipedia as a source of online attention**

Wikipedia is a free encyclopedia written collaboratively by a myriad of users through an open process. Volunteers governed by a social system build content by creating and editing encyclopedia articles (Oeberst et al., 2014)[1]. Their content has outstanding reach and visibility, as evidenced by the 230 million daily views received by its English edition in 2022[2] as the only non-commercial website among the ten most popular websites worldwide[3]. Also, the active Wikipedias[4] in 321 languages allow investigation of the different social and cultural features among regions (Miquel-Ribé & Laniado, 2018; Roy et al., 2022), measure public interest on diverse topics (Roll et al., 2016), and enable production of monitoring and forecasting systems (Generous et al., 2014; Mittermeier et al., 2021).

Despite some early concerns about its reliability as a source of information (Gorman, 2007)[5], Wikipedia is noted for its great epistemic benefits which offset its potential lack of accuracy. Fallis (2008) points out that these benefits relate to *speed* (how quickly knowledge can be acquired), *power* (how much knowledge can be acquired), and *fecundity* (how many people can acquire that knowledge). Thus, university Wikipedia pages represent a point of aggregation allowing the general public to gather information about particular universities, and can be considered as a reflection of one form of online attention. Other more specific measures of attention include those related to users' co-creation activities: the number of times a Wikipedia article has been edited, the number of users who have participated in its discussion, and the number of references used (Arroyo-Machado et al., 2022).

Wikipedia is used frequently in altmetrics studies (Colavizza, 2020; Nielsen, 2007; Yang & Colavizza, 2022). There are numerous reference-based studies exploring the coverage and characteristics of publications cited in Wikipedia (Kousha & Thelwall, 2017; Lewoniewski et al., 2017; Nicholson et al., 2021; Teplitskiy et al., 2017; Torres-Salinas et al., 2018), and several works collect these citations which are used to generate and publish datasets (Singh et al., 2020; Zagorova et al., 2022), or adapt classical scientometrics techniques such as co-citation networks (Arroyo-Machado et al., 2020; Didegah & Thelwall, 2018). Some authors such as Maggio et al. (2020) study engagement patterns. These authors found that frequently Wikipedia readers check the sources of medical articles although they rarely access them confirming the findings in Piccardi et al. (2020) that the number of users accessing cited publications is very small.

**Measuring universities' reputation and social media attention**

Scientific reputation and social media attention have been operationalized primarily using university rankings. Webometric literature through link analysis and web log files shows that the position in the academic ranking is correlated with the interest and attention universities gather online (Aguillo et al., 2006; Vaughan & Romero-Frías, 2014). These analyses have been extended to include Wikipedia and reveal a connection between academic rankings and the university's web presence and attention, measured through interlinks among Wikipedia articles (Eom et al., 2013; Zhirov et al., 2010) and page views (Katz & Rokach, 2017).

A commonly used approach is use of a social network lens to estimate the relative importance of Wikipedia university pages, and algorithms such as PageRank or HITS (Eom et al., 2013; Zhirov et al., 2010). This approach has been replicated and extended and is considered better able to capture universities' economic and cultural impact compared to bibliometric-based rankings (Li et al., 2019). One such example is the Wikipedia Ranking of World Universities (WRWU) which is published in 24 languages (Coquidé et al., 2019; Lages et al., 2016). Other rankings combine network analysis with the scientific capital of their faculty members and alumni in an attempt to more accurately reflect popular perceptions of reputation (Katz & Rokach, 2017). Regardless of the method employed (see Table 1), there are considerable overlaps if Wikipedia-based rankings are compared to rankings based on research performance indicators (e.g. the Academic Ranking of World Universities, ARWU).

**Table 1**

*Similarities between Wikipedia-based university rankings and rankings based on research performance*

| Paper | Method | Rankings | Overlap/Correlation |
|---|---|---|---|
| Zhirov et al. (2010) | Network Analysis | Top 100 ARWU2009 | Overlap 70% |
| Eom et al. (2013) | Network Analysis | Top 10 ARWU03/05/07/09/11 | Overlap 80% |
| Lages et al. (2016) | Network Analysis | Top 100 ARWU2013 | Overlap 62% |
| Katz & Rokach (2017) | Network Analysis + Indicators | 390 ARWU2011 | Spearman's ρ = 0.53 |
| | | 390 THE2011 | Spearman's ρ = 0.47 |
| Coquidé et al. (2019) | Network Analysis | Top 100 ARWU2017 | Overlap 60% |
| Li et al. (2019) | Network Analysis + Indicators | Top 114 THE2015 | Pearson's ρ > 0.6 |
| | | Top 114 QS2015 | Spearman's ρ > 0.5 |
| Li et al. (2020) | Network Analysis | Top 100 THE | Overlap 55% |
| | | Top 100 ARWU | Overlap 54% |
| | | Top 100 QS | Overlap 51% |

**Zooming in on the relationship between reputation and attention**

The above studies reveal a connection between universities' scientific reputation and Wikipedia attention but the specific factors related to research performance and their interaction with Wikipedia attention needs more investigation. All the studies cited above assume that university reputation is based on the university's overall position in the rankings which are based on composite or weighted indicators. However, composite indicators lack theoretical justification due to the arbitrariness in the choice of weights and the possible interdependence of components defining the composite indicator (Glänzel & Debckere, 2009), making the aggregate picture rankings provide likely inaccurate of the association between research performance and the attention paid to a university. A more granular approach is need to unpack this relationship.

Performance measures related to productivity, scientific impact, and international collaboration (see for a review, Tahamtan et al., 2016) do not necessarily translate into social media attention as they do in the case of scientific reputation. Thus, their effect needs to be tested individually and accounting for contextual factors (location, language, university orientation) which might affect the relationship. In relation to social media attention, a more granular approach implies the need to bring the geographical scope of the attention (i.e. international vs. local) into play and examine whether research performance indicators have a similar effect across all universities' on Wikipedia. Put differently, factors related to a university's scientific reputation might be relevant regardless of the source of the page views but their relative importance might vary. Based on this discussion, we address the following research questions:

1. Is a university's scientific reputation associated to the attention it receives on Wikipedia? Is the effect the same across all regions?
2. Does the strength of the association vary depending on the attention scope (international or local) considered?

## Methodology

**Data collection and indicators**

Our data collection involved the integration of data from two different sources. First, the Leiden Ranking (Waltman et al., 2012) was used to collect bibliometric indicators[6] of university research performance. The Leiden Ranking is based on an enriched in-house database version of the Web of Science (WoS). The 2021 edition includes 1,225 universities from 69 countries with a strong research orientation and good scientific impact (van Eck, 2021). Inclusion in the 2021 edition requires the university to have published a minimum 800 WoS publications (articles and reviews, hereafter papers) in the period 2016-2019. Use of this ranking is based on its reliability for evaluation purposes (Vernon et al., 2018) and its diverse indicators.

We considered several dimensions of research performance which constitute university symbolic capital: productivity, collaboration, scientific impact, thematic specialization, and gender. Productivity is measured as total number of WoS papers in the period of analysis (*P*) and proportion of papers included in the Dimensions database and not indexed in WoS (*Dimensions WoS*). We measure collaboration based on number of co-authored papers involving more than one organization (*P collab*), and at least one industry partner (*P industry collab*), and authors affiliated to organizations in different countries (*P international collab*). We employed size-dependent and independent indicators to measure scientific impact in terms of total number of citations to papers published by a particular university (*TCS*), total number of citations normalized by field and publication year (*TNCS*) and percentage of papers included in the top 10% most cited (*P top 10*). For thematic specialization, we use the five major scientific fields in the Leiden Ranking[7] to define universities as specialized in biomedicine, health sciences, and life and earth sciences (*Topic specialization bio*) or in mathematics, computer science, physics and engineering (*Topic specialization eng*) if the proportion of the papers in the areas identified exceeds 50% of their total output in WoS[8]. Finally, we take account of gender based on number of female authorships as a proportion of a university's number of male and female authorships *(Pa f mf)*.

Second, for each university included in the Leiden Ranking, we retrieved the corresponding English language Wikipedia page. This resulted in the matching of 1,220 universities from the Leiden Ranking (99.6% of the total) with their English Wikipedia page. Next, we identified the respective university articles in other language editions which resulted in a total of 27,374 articles in 225 Wikipedias. For each Wikipedia article, we collected metadata on its content i.e. number of characters (*Characters*), words (*Words*), sections (*Sections*), references (*References*), and language versions (*Language links*). We also considered users' activity measured as number of editions (*Editions*) and editors (*Editors*). Attention was estimated based on university universities' page views between July 2015 (the earliest data available) to May 2022. We conducted our data collection on May 30, 2022 using the Wikimedia REST API[9] and the XTools API Page[10]. We distinguished Wikipedia page views based on geographical location to account for different types of attention: (i) the total number of page views received by the university in the different language editions (*Total views)*; (ii) total number of page views of the English edition of Wikipedia to proxy for international attention (*International views*); and (iii) the total page views in all the official languages and dialects in each country (*Local views*)[11].

We took account also of geographical and linguistic factors to obtain a finer grained view of the universities' profiles. We included a dummy variable which takes the value 1 for universities located in the Anglosphere (*Anglo country)*—Australia, Canada, New Zealand, United Kingdom and United States—, and a categorical variable to account for the region (*Continent)* of each university—Africa, Europe, North America, Oceania, South America, Asia. We also considered number of years since university foundation (*Age*). Table 2 presents the indicators used. The scripts employed to retrieve and process the data are available from Jupyter's Notebooks at GitHub (https://doi.org/10.5281/zenodo.8092586) and the dataset can be accessed at Zenodo (https://doi.org/10.5281/zenodo.8092479).

**Table 2**

*Summary of the indicators employed*

| Dimension | Indicator | Measure description |
|---|---|---|
| **Scientific performance** | | |
| **Productivity** | *P* | University's total no. of papers (full count) |
| | *Dimensions WoS* | Percentage of the university's Dimensions papers not indexed in WoS |
| **Collaboration** | *P collab* | Number of papers co-authored with one or more other organizations |
| | *P industry collab* | Number of papers co-authored with one or more industry organizations |
| | *P int collab* | Number of papers involving authors from 2 or more countries |
| **Scientific impact** | *TCS* | Total no. of citations |
| | *TNCS* | Total no. of citations normalized by field and publication year |
| | *P top 10%* | Number of papers that, compared with others papers in the same field and year, belong to the top 10% most frequently cited papers |
| **Thematic specialization** | *Topic specialization bio* | Dummy = 1 if at least 50% of the university's papers are in biomedical and health sciences/life and earth sciences |
| | *Topic specialization eng* | Dummy = 1 if at least 50% of the university's papers are in mathematics, computer science/physical sciences and engineering |
| **Gender** | *Pa f mf* | The proportion of female authors |
| **Wikipedia** | | |
| **Content** | *Characters\** | Total no. of characters in the university's Wikipedia article |
| | *Words\** | Total no. of words in the university's Wikipedia article |
| | *Sections\** | Total no. of sections in the university's Wikipedia article |
| **Edits** | *Edits\** | Total no. of edits made to the university's Wikipedia article |
| | *Editors\** | Total no. of editors contributing to the university's Wikipedia article |
| **Views** | *Total views* | Total no. of views of the university's Wikipedia article |
| | *International views* | Total no. of views of the university's English language Wikipedia article |
| | *Local views* | Total no. of views of the university's total Wikipedia articles in languages and dialects recognized in the university's country of location |
| **Internalization** | *Language links* | Total no. of links to other language editions of the university's Wikipedia article |
| **References** | *References\** | Total no. of references in the university's Wikipedia article |
| | *Unique references\** | Total no. of unique references in the university's Wikipedia article |
| **University** | | |
| **Location** | *Anglo country* | Dummy = 1 if the university is located in Australia, Canada, New Zealand, United Kingdom, or United States |
| | *Continent* | Categorical variable: Africa, Europe, North America, Oceania, South America, Asia [ref. category] |
| **Foundation** | *Age* | Number of years since university was established |

\* *Note.* These totals are the sum of all university articles in the different language editions of Wikipedia

**Exploratory analysis**

We provide a descriptive analysis by country and language and use Spearman's rank correlations to understand the overall relationship between a university's Wikipedia page views and its measures of research performance. We examine the differences between the local and international attention received by the university, and account also for the eight private universities in the Ivy League given their status as among the world's most prestigious academic institutions. This distinction allows us to assess whether this recognition has an influence on local attention in non-Anglo-Saxon regions.

**Regression analysis**

We test whether performance measures related to productivity (*P*), international collaboration (*P int collab*), and scientific impact (*P top 10*) are associated to the attention universities receive on Wikipedia and if the same relationship holds for other aspects than university economic and social capital. To reduce collinearity between the performance measures, we categorize international collaboration and productivity by terciles: low ($\leq P_{33}$)—ref. category—, intermediate ($>P_{33}$ and $\leq P_{66}$), and high ($> P_{66}$). Our dependent variable is total number of Wikipedia pages views (*Total views*). We ran additional models focusing on the number of *international views* and *local views* to test a potential local component affecting attention. Since our dependent variables are count variables, we used Negative Binomial estimations. Evidence of overdispersion (e.g., *total views*: deviance goodness-of-fit = 598,000,000, p = 0.000) excluded use of a Poisson model. The models include *Anglo country*, *Continent*, *Age*, *Topic specialisation bio, Topic specialisation eng*, *Pa f mf*, and *Dimension WoS* to control for geographical and language determinants. All variables are standardized so that the size of the effects is based on a one-unit change in the dependent variable.

<div align="center">

**Results**

</div>

**General overview**

The distribution of universities in the Leiden Ranking by country is skewed. China is ranked first with 217 universities followed by the United States with 200 and United Kingdom with 61 universities. There are 21 countries with at least 10 universities, representing a total of 1,044 universities (86% of the total analyzed).

Table 3 summarizes the results for scientific reputation and Wikipedia attention for the 21 countries with the highest number of universities in the Leiden Ranking. The Netherlands (12 universities) and Sweden (13) have high scientific impact (in terms of number of highly cited papers and total number of normalized citations) and high levels of international collaboration. The Anglo-Saxon countries—the United States (200 universities), the United Kingdom (61 universities) and Australia (32 universities) show high research performance while the universities in India (38), Iran (36), and Turkey (32) achieve comparatively lower values across the different dimensions of research performance.

The Wikipedia metrics reveal similar patterns. Attention is notable among the Anglosphere countries based on Wikipedia page views. Although China has the highest number of universities in the Leiden Ranking, it receives lower levels of national and international attention. The levels of national attention are likely due to China's having its own collaborative encyclopedia, Baidu Baike, and the fact that Wikipedia is blocked in mainland China[12]. Beyond the Anglosphere, differences in attention are less pronounced than comparisons based on scientific reputation.

**Table 3**

*Average values for universities' Wikipedia and research performance indicators by country (n = 1,220)*

| Country | University | | Research performance | | | Wikipedia | | | |
| | No. | Age | P | P int collab | P top 10% | Language links | Local views | International views | Edits |
|---|---|---|---|---|---|---|---|---|---|
| China | 217 | 89.4 | 8015.9 | 2254.1 | 917 | 8.2 | 95825.9 | 93236.2 | 799.6 |
| United States | 200 | 147.3 | 10694.9 | 4621.6 | 1796.4 | 31.2 | 1979702.4 | 1979702.4 | 4906.3 |
| United Kingdom | 61 | 161.8 | 9448.3 | 6180.3 | 1679.3 | 37.5 | 1244018.8 | 1242904.7 | 4350.6 |
| Germany | 54 | 232 | 8322.2 | 4908.8 | 1217.3 | 36.5 | 320047.9 | 291413.9 | 2932.4 |
| Japan | 54 | 104.6 | 5765.9 | 1974.7 | 494.6 | 16.4 | 756578.5 | 130824.8 | 1633.9 |
| South Korea | 46 | 86.6 | 6492.7 | 1956.2 | 532 | 12.6 | 121928 | 248131.2 | 1549.7 |
| Italy | 42 | 323.1 | 7596.4 | 4058.7 | 980.5 | 25.5 | 262138.7 | 178701.6 | 1714.4 |
| Spain | 42 | 219.3 | 5305.1 | 2921.9 | 608.9 | 21.7 | 245805.1 | 116310.5 | 1346.1 |
| India | 38 | 74.1 | 3349.7 | 910.6 | 278.2 | 15.1 | 788956.2 | 745249.5 | 1822.2 |
| Iran | 36 | 61.5 | 3797.6 | 1066.3 | 333.1 | 8.1 | 242611.4 | 60166.7 | 985.5 |
| Australia | 32 | 73.5 | 10643.6 | 6275.8 | 1631.8 | 24.7 | 563584.3 | 563584.3 | 2430.8 |
| Turkey | 32 | 78.8 | 2702.6 | 819.3 | 185.8 | 14.5 | 201000.8 | 114498.3 | 1219.9 |
| Brazil | 31 | 71.4 | 6191.3 | 2573.9 | 501.1 | 17.8 | 181464.9 | 48964.7 | 1165 |
| Poland | 31 | 116.7 | 3139.3 | 1205.1 | 244.1 | 18.3 | 171856.4 | 80605.7 | 1125.6 |
| Canada | 30 | 120.9 | 10716.4 | 6024.1 | 1538.9 | 30.9 | 1079240.8 | 995113.6 | 3521.5 |
| France | 30 | 272.9 | 10101.3 | 6278.6 | 1442.6 | 27.3 | 147720.5 | 166818.3 | 1291.4 |
| Taiwan | 21 | 78.7 | 5128.4 | 1754.3 | 408.2 | 13.1 | 570572.1 | 99502.3 | 2151.9 |
| The Netherlands | 13 | 194.9 | 15213.8 | 9608.5 | 2604.1 | 39.3 | 131776.4 | 447745.3 | 2715.2 |
| Austria | 12 | 284.9 | 4813.4 | 3416.4 | 685.8 | 24.9 | 141606.1 | 190930.5 | 1467.7 |
| Sweden | 12 | 163.7 | 11142.8 | 7558.3 | 1651.8 | 38.6 | 132864.3 | 330566.3 | 2410.5 |
| Russia | 10 | 138.2 | 5009.7 | 3017.4 | 400.8 | 35.5 | 772638.6 | 306399.3 | 2995.3 |

Figure 1 depicts the Spearman correlation matrix for research performance and Wikipedia indicators for the 1,220 universities analyzed. Overall, there are moderate to strong positive correlations between both worlds, the scientific and social media. For instance, the correlations between raw and normalized citations and Wikipedia indicators range between 0.5 and 0.6. Note that productivity ($P$) is more strongly correlated to *international* (0.49) than *local* (0.36) *views*, while the correlation to *local views* is stronger if publications not indexed in the WoS (*Dimension WoS*) are taken as a reference (0.24 vs 0.41). This suggests that the scope of the university's production matters for the type of Wikipedia attention. However, degree of specialization in mathematics, computer science, physics, and engineering (*Topic specisalization eng*) is weakly but negatively correlated with the Wikipedia indicators, suggesting that technical universities receive less online Wikipedia attention. Figure S1 depicts the correlations with Wikipedia indicators.

**Figure 1**

*Spearman correlations between Wikipedia and research performance indicators (n = 1,220)*

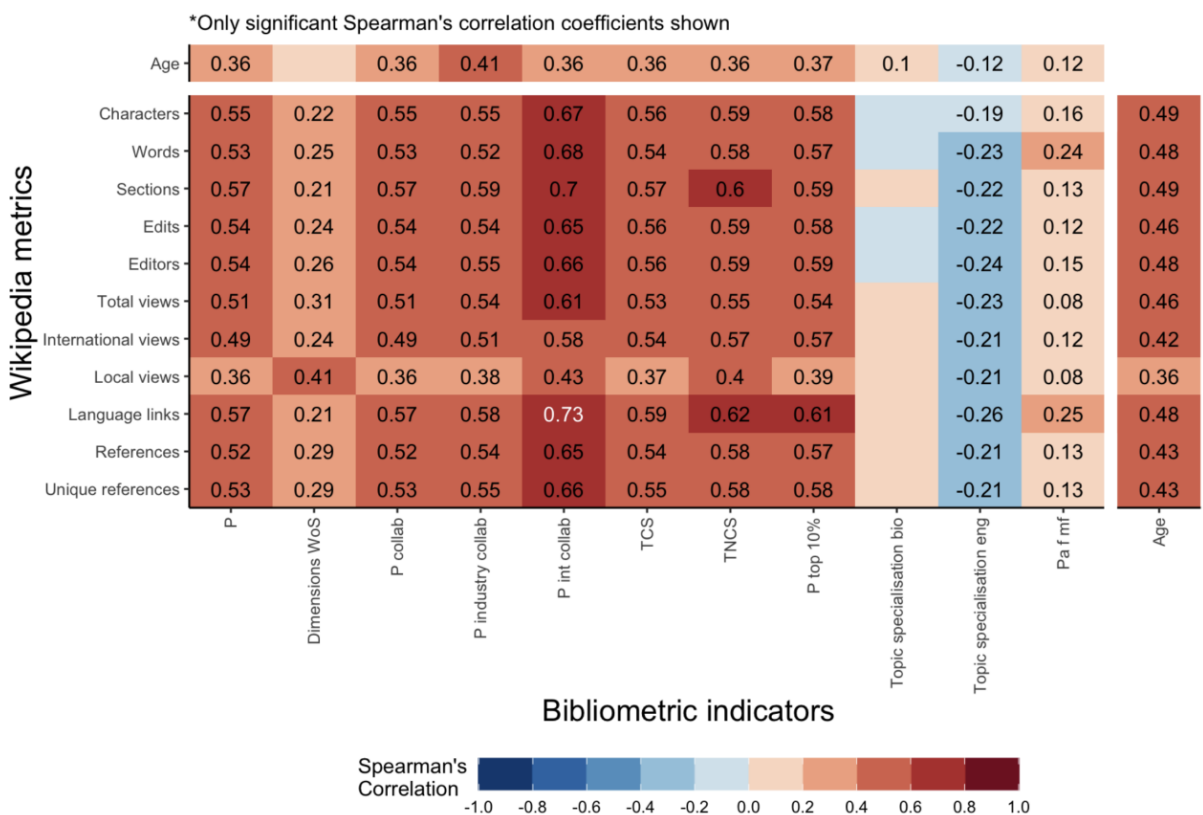

**Language dimension**

Plotting international and local views by language edition provides additional evidence on the scope of Wikipedia attention (see Figure 2). It is clear that regardless of the language, Ivy League universities (green dots) attract the highest number of international and local views. Beyond this elite group, universities located in countries with languages different from the official one/s for that language area receive considerably less attention (blue dots). However, for areas with "strong" languages such as Spanish (es), Persian (fa), and Japanese (ja), the universities from those countries (red dots) receive high levels of local attention which greatly exceeds the international attention they attract. This pattern becomes less clear cut due to certain country specifics and particularly for the group of universities in the German (ge) language area. In the Tyrol region of Italy, German is an official language which results in some Italian universities being labeled German (red dots). Nevertheless, the scatterplots suggest a potential local component related to universities' Wikipedia attention which is deserving of closer scrutiny.

**Figure 2**

*Distribution of local and international universities' Wikipedia page views by language and country*

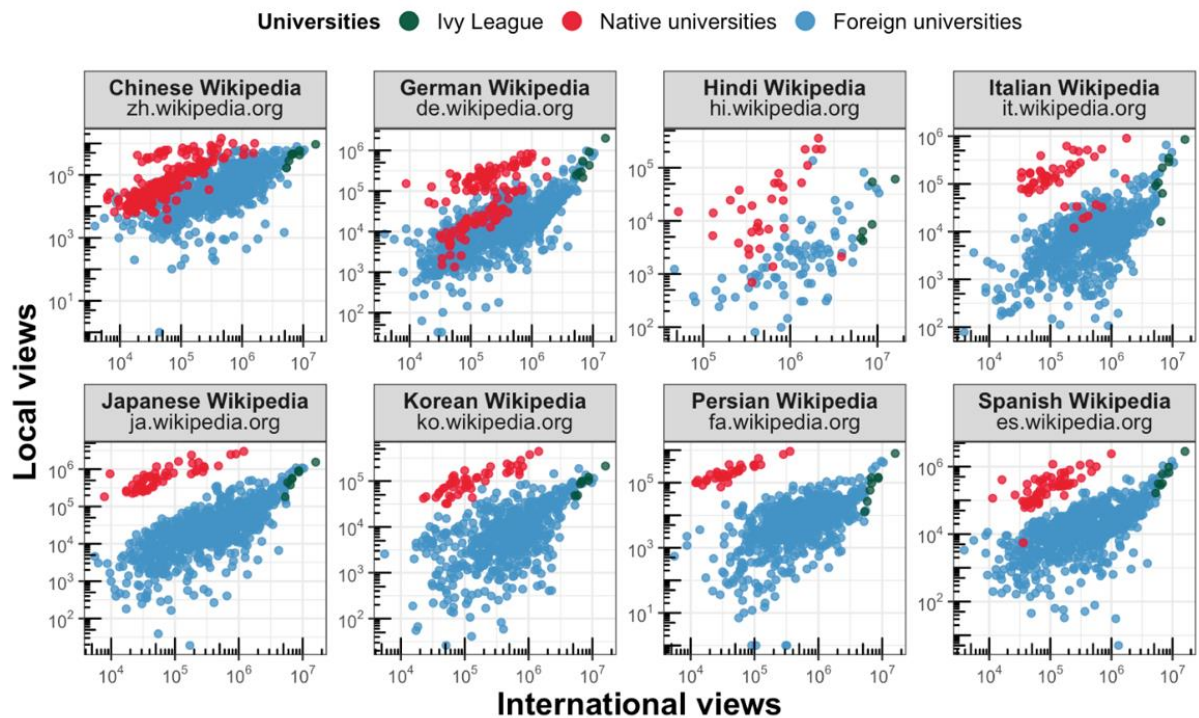

*Note.* The Ivy League is an elite group of American universities including Brown University, Columbia University, Cornell University, Dartmouth College, Harvard University, University of Pennsylvania, Princeton University, and Yale University.

**Regression results**

We employed Negative Binomial estimations to test the relationship between universities' scientific reputation and the online attention they attract. To ease comparison between the models, we interpret the coefficients in terms of incidence rate ratios (IRR). Our dependent variable is number of universities' Wikipedia page views which by definition is a rate.

Table 4 reports the main results for total number of Wikipedia page views. The baseline model includes the geographical and language control variables and the indicators for university profile. Note that the estimated ratios for percentage of female authors and specialization in math, computer science, physics, and engineering are negative and significant while location in an Anglo-Saxon country and North America are positive and significant. Model 1 adds productivity. We see that medium (z = 10.79, p = 0.000) and high levels (z = 28.05, p = 0.000) of productivity have a significant effect on university attention at the 1% level. The IRR indicates that when comparing highly to less productive universities given the other measures in the model are held constant, the IRR for total Wikipedia page views is 3.68 times greater. Model 2 includes international collaboration. Again, the effect is highly significant and positive, and total page views increase by 1.84 for intermediate levels of international collaboration and 4.49 for high levels of international collaboration. Model 3 shows that the number of top publications has a positive and highly significant effect (z = 7.42, p = 0.000) and increases Wikipedia attention by 1.78. Model 4 is the full model. We find support for our assumption that scientific reputation is associated to greater Wikipedia attention, due to the effect of highly cited papers (IRR = 1.39, z = 8.77, p = 0.000) and medium (IRR = 1.67, z = 5.60, p = 0.000) and high (IRR = 2.47, z = 4.13, p = 0.000) levels of international collaboration which are positive and highly significant. In the full model, the effect of productivity disappears due to the interdependence with the other two explanatory variables.

**Table 4**

*Results for Negative Binomial regression. Dependent variable: Total views (N = 1,220)*

| | M0: Baseline model | M1: Productivity | M2: International collab. | M3: P top 10% | M4: Full model |
|---|---|---|---|---|---|
| | IRR (SE) | IRR (SE) | IRR (SE) | IRR (SE) | IRR (SE) |
| P top 10% | | | | 1.78*** (0.14) | 1.39*** (0.05) |
| *P int collab.* | | | | | |
| Medium int. collab. | | | 1.84*** (0.13) | | 1.67*** (0.15) |
| High int. collab. | | | 4.49*** (0.43) | | 2.47*** (0.54) |
| *P* | | | | | |
| Medium productivity | | 1.59*** (0.07) | | | 1.06 (0.04) |
| High productivity | | 3.68*** (0.17) | | | 1.10 (0.15) |
| Pa f mf | 0.82*** (0.05) | 0.92 (0.05) | 0.92*** (0.03) | 0.91*** (0.03) | 0.93*** (0.02) |
| Dimensions WoS | 1.28* (0.17) | 1.50*** (0.13) | 1.52*** (0.11) | 1.50*** (0.11) | 1.57*** (0.09) |
| Age | 1.49*** (0.14) | 1.29*** (0.09) | 1.30*** (0.08) | 1.28*** (0.07) | 1.24*** (0.07) |
| Specialization bio | 1.12 (0.30) | 0.94 (0.15) | 0.10 (0.13) | 0.78*** (0.07) | 0.83*** (0.06) |
| Specialization eng | 0.79* (0.09) | 0.95 (0.12) | 0.95 (0.10) | 0.86 (0.12) | 0.91 (0.12) |
| Anglo-saxon country | 2.64*** (0.27) | 2.65*** (0.30) | 2.32*** (0.31) | 1.91*** (0.24) | 1.93*** (0.26) |
| *Continent* | | | | | |
| Africa | 0.90 (0.24) | 0.76 (0.14) | 0.52*** (0.09) | 0.77 (0.12) | 0.56*** (0.10) |
| Europe | 1.05 (0.12) | 0.86 (0.08) | 0.656*** (0.06) | 0.90 (0.08) | 0.71*** (0.06) |
| North America | 1.85*** (0.08) | 1.33*** (0.11) | 1.32*** (0.13) | 1.45*** (0.19) | 1.32** (0.16) |
| Oceania | 0.77*** (0.07) | 0.51*** (0.06) | 0.45*** (0.06) | 0.66*** (0.09) | 0.51*** (0.08) |
| South America | 0.79 (0.28) | 0.53** (0.13) | 0.42*** (0.09) | 0.58*** (0.12) | 0.42*** (0.07) |
| Constant | 473309*** (35612.70) | 273848.50*** (15518.51) | 275557.10*** (10581.83) | 508805.70*** (34176.03) | 327279.30*** (16316.04) |
| Cragg & Uhler's $R^2$ | 0.506 | 0.506 | 0.680 | 0.669 | 0.717 |

*Note.* Robust standard errors are clustered by continent and reported in parentheses; *** $p<0.01$, ** $p<0.05$, * $p<0.1$

To deepen the relationship between scientific reputation and online attention, we distinguish between local (Table S1) and international (Table S2) Wikipedia page views.

Overall, the previous patterns persist: high levels of productivity, international collaboration, and highly cited papers are positively and significantly associated to the number of Wikipedia local views. If we consider only Wikipedia international page views the results do not change. However, in terms of the magnitude of the effects for total, international and local views, research performance clearly has the smallest effect on local attention (see Table 5).

**Table 5**

*IRR comparison for the Wikipedia total, international, and local page views (N = 1,220)*

| Pages views | Productivity (*P*) | | International collab. (*P int collab*) | | Highly cited papers (*P top 10%*) |
|---|---|---|---|---|---|
| | Medium | High | Medium | High | |
| | IRR | IRR | IRR | IRR | IRR |
| Total | 1.59 | 3.68 | 1.84 | 4.49 | 1.78 |
| International | 1.58 | 3.86 | 1.86 | 4.87 | 1.86 |
| Local | 1.40 | 2.59 | 1.60 | 3.05 | 1.53 |

Finally, we test for potential curvilinear effects to check the extent to which a high level of research performance linearly increases universities' attention on Wikipedia. We found an inverted U-shape relationship between the three main explanatory variables analyzed and total number of Wikipedia page views (see Table S3). The linear terms are positive and significant and the squared terms are negative and significant for productivity (IRR = 2.05, z = 13.48, p = 0.000; IRR = 0.95, z = -4.33, p = 0.000), international collaboration (IRR = 2.213, z = 9.430, p = 0.000; IRR = 0.939, z = -4.500, p = 0.000), and highly cited papers (IRR = 2.048, z = 11.140, p = 0.000; IRR = 0.959, z = -5.280, p = 0.000), suggesting the presence of diminishing returns from strong research performance in terms of Wikipedia page views. Figure 3 plots the predicted marginal effects. These results do not change in the case of international and local views.

**Figure 3**

*Effect of productivity, international collaboration and highly cited papers on the total number of*
*Wikipedia pages views*

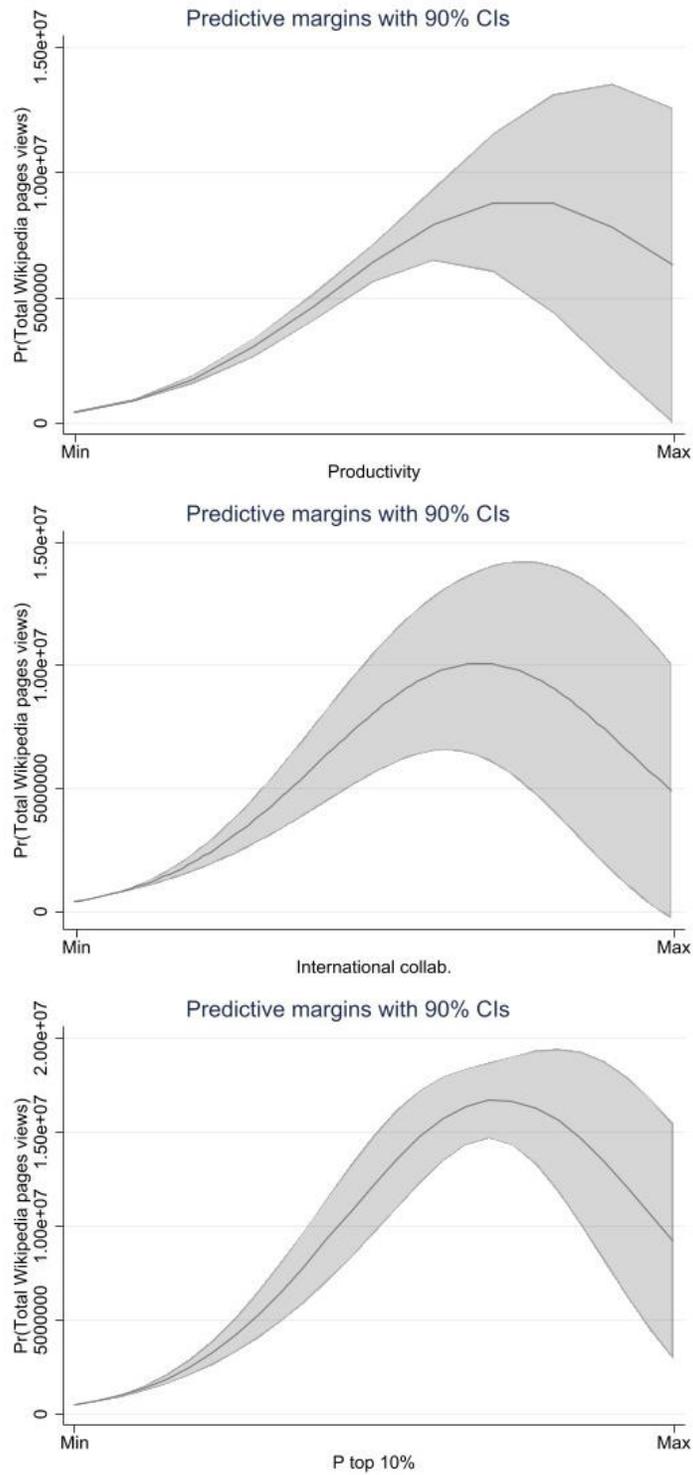

**Discussion**

In this study, we examined the extent to which a university's scientific reputation is related to the social media attention it receives on Wikipedia. We relied on a range of indicators from the Leiden Ranking to measure research performance and page views of university Wikipedia articles to proxy for online attention. Our study departs from previous web-based studies which rely on weighting schemes and composite indicators taken from university rankings and which lack theoretical and empirical grounding. We adopted a more granular approach in line with the sensitivity of composite scores to small changes in the weights assigned to each individual measure and the need to examine the effect of each performance measure individually. This approach is in line also with use of multi-faceted data to explore university performance patterns rather than relying only on "final" or "perfect" operationalization of academic excellence (Moed, 2017).

Our exploratory analysis reveals that science and social media are positively correlated, suggesting that university reputation—measured by bibliometric indicators—is aligned to some extent to the online interest in the university among the general public. We also found differences in the amount of attention received based on the university's research specialization; those more multidisciplinary and internationally oriented receive more international attention. The level of Wikipedia attention varies according to the language edition, demonstrated by comparison of Wikipedia page views in different language editions. The regression models show that productivity, international collaboration, and highly cited papers are associated with greater attention on Wikipedia. This effect cuts across regions and linguistic areas and whether attention is international or local—although the effect is less pronounced in the case of local attention. However, we found evidence also of a curvilinear effect for the three measures of scientific performance which suggests that translation of university scientific symbolic capital into Wikipedia attention has a positive relationship but at a decreasing rates (i.e. it does not increase linearly it follows and inverted-U shape). This suggests that other factors beyond the university's scientific reputation might be driving the attention on Wikipedia, and particularly attention to Wikipedia pages in local languages.

Future research could include other sources of variation in universities' performance and activities to allow better identification of which factors related to universities' symbolic capital are driving the online attention the universities receive. The main challenge in this regard is the lack of systematic large-scale data on other university activities such as regional engagement, knowledge transfer, economic outcomes, and public health impact (Vernon et al., 2018). The U-Multirank provides detailed data on some of these dimensions but relies on self-reported information which results in poor data completeness (>60% of missing data for the universities included in the Leiden Ranking[13]).

Our work has some practical limitations. First, the bibliometric indicators from the Leiden Ranking are based on WoS publications with the result that certain universities, knowledge domains, and specialties with a stronger local orientation are underrepresented; this limitation applies particularly to arts and humanities subjects (Petr et al., 2021). Second, we capture local attention on Wikipedia not at the country level but at the linguistic regional level which can include more than one country. Therefore, this does not provide an accurate portrayal of universities' local page views at the country level (e.g. Spanish Wikipedia includes all Spanish-speaking countries). Third, the existence of countries which have multiple official languages and dialects hinders analysis of the local component, as shown in the case of Italy for instance.

### Concluding remarks

Our study contributes to a growing stream of work which conceives altmetrics as analytical tools to investigate science-society interactions. Our analytical approach moves away from the still dominant one-dimensional views provided by web-based studies which use data from academic rankings and advocate for more granular analyses. We show that universities' scientific performance is associated strongly with the online attention they receive on Wikipedia, though differences exist when distinguishing between international and local views. However, we show that this relationship is not linear and found evidence of diminishing returns (inverted U-shape) which highlights the need to extend analyses of the factors driving online attention beyond research performance measures.

### Acknowledgements

Wenceslao Arroyo-Machado is supported by a FPU Grant (FPU18/05835) from the Spanish Ministry of Universities and acknowledges funding from a project by MCIN (PID2019-109127RB-I00/SRA/10.13039/501100011033). Adrián A. Díaz-Faes acknowledges research project PID2020-112837RJ-I00 funded by MCIN/AEI/ 10.13039/501100011033. Rodrigo Costas is partially funded by the South African DSI-NRF Centre of Excellence in Scientometrics and Science, Technology and Innovation Policy (SciSTIP). A draft version of this paper was presented at the 26th STI Conference (Granada, 2022).

**Author contributions**

WAM – Data curation; Formal Analysis; Investigation; Resources; Software; Visualization; Writing – original draft

AADF – Conceptualization; Methodology; Formal Analysis; Funding acquisition; Investigation; Software; Writing – original draft; Writing – review & editing

EHV – Supervision; Writing – review & editing

RC – Conceptualization; Methodology; Project administration; Supervision; Validation; Writing – review & editing

**Endnotes**

[1]  Yun et al. (2019) show that Wikipedia faces challenges related to unequal participation and some topics being dominated by small groups of super-editors which might jeopardize its aim of democratizing knowledge in the long run.

[2]  https://pageviews.wmcloud.org/siteviews/?platform=all-access&source=pageviews&agent=user&start=2022-01-01&end=2022-12-31&sites=en.wikipedia.org (Accessed on 3 May 2023)

[3]  https://en.wikipedia.org/wiki/List_of_most_visited_websites (Accessed on 3 May 2023)

[4]  https://meta.wikimedia.org/wiki/List_of_Wikipedias (Accessed on 3 May 2023)

[5]  Inaccuracies, errors, and omissions in Wikipedia are comparable to those found in traditional encyclopedias (Giles, 2005). Among contributors, highly committed participants and anonymous users who contribute show high reliability in the contributions (Anthony et al., 2009).

[6]  https://www.leidenranking.com/information/indicators (Accessed on 3 May 2023)

[7]  Biomedical and health sciences; life and earth sciences; mathematics and computer science; physical sciences and engineering; and social sciences and humanities.

[8]  Only two universities are specialized in social sciences and humanities: the London School of Economics and Political Science and Tilburg University. This is due largely to the well-known limited coverage by the Web of Science of these two areas.

[9]  https://www.mediawiki.org/wiki/Wikimedia_REST_API (Accessed on 3 May 2023)

[10]  https://www.mediawiki.org/wiki/XTools/API/Page (Accessed on 3 May 2023)

[11]  For the Anglo-Saxon countries, international and local views mostly overlap; however, except for the United Kingdom (Cornish, Irish, Scottish Gaelic, and Welsh) and Canada (French), Wikipedia university pages are available in languages and dialects other than English.

12  https://en.wikipedia.org/wiki/List_of_websites_blocked_in_mainland_China
    (Accessed on 3 May 2023)

13  Using a dataset provided by U-Multirank (data from 2021), we matched 83% of the
    universities from the Leiden Ranking. We found a large number of missing values for
    most indicators not based on research outputs. For instance, bachelor's graduation
    rates were available only for 34% of these universities, and student mobility was
    available for only 25%.

# Appendix

## Figure S1

*Correlations between Wikipedia metrics. Only significant Spearman's correlation coefficients shown (p < 0.05)*

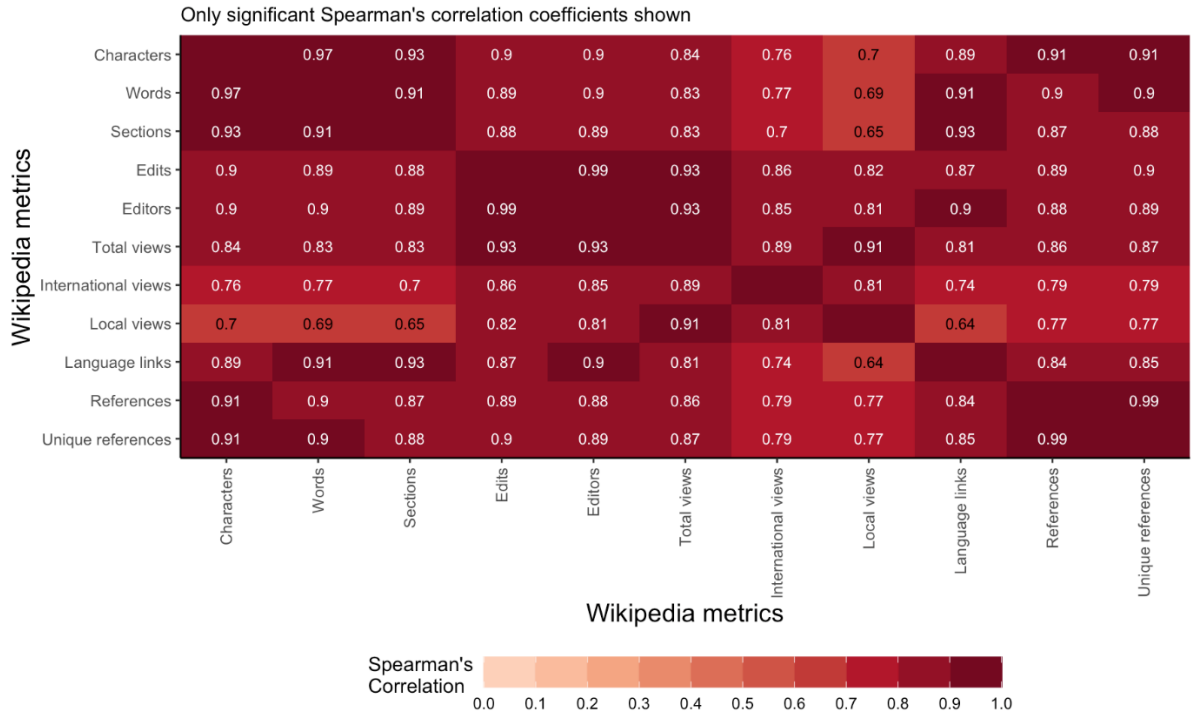

**Table S1**

*Results for Negative Binomial regression. Dependent variable: Local views (N = 1,220)*

| | M0: Baseline Model | M1: Productivity | M2: International collab. | M3: P top 10% | M4: Full model |
|---|---|---|---|---|---|
| | IRR (SE) | IRR (SE) | IRR (SE) | IRR (SE) | IRR (SE) |
| P top 10% | | | | 1.53*** | 1.31*** |
| | | | | (0.04) | (0.04) |
| *P int collab.* | | | | | |
| Medium int. collab. | | | 1.60*** | | 1.48*** |
| | | | (0.05) | | (0.10) |
| High int. collab. | | | 3.05*** | | 1.92*** |
| | | | (0.15) | | (0.26) |
| *P* | | | | | |
| Medium productivity | | 1.40*** | | | 1.04 |
| | | (0.03) | | | (0.05) |
| High productivity | | 2.59*** | | | 1.06 |
| | | (0.11) | | | (0.10) |
| Pa f mf | 0.88*** | 0.95 | 0.94*** | 0.94** | 0.96*** |
| | (0.04) | (0.03) | (0.02) | (0.02) | (0.01) |
| Dimensions WoS | 1.50** | 1.66*** | 1.67*** | 1.66*** | 1.71*** |
| | (0.22) | (0.19) | (0.17) | (0.17) | (0.16) |
| Age | 1.34*** | 1.21*** | 1.21*** | 1.19*** | 1.17*** |
| | (0.09) | (0.06) | (0.05) | (0.04) | (0.04) |
| Specialization bio | 1.08 | 0.99 | 1.04 | 0.86*** | 0.90** |
| | (0.19) | (0.08) | (0.06) | (0.03) | (0.04) |
| Specialization eng | 0.97 | 1.11 | 1.13* | 1.05 | 1.09 |
| | (0.07) | (0.09) | (0.08) | (0.08) | (0.09) |
| Anglo-saxon country | 4.43*** | 4.30*** | 3.85*** | 3.24*** | 3.27*** |
| | (0.90) | (0.85) | (0.88) | (0.71) | (0.74) |
| *Continent* | | | | | |
| Africa | 1.00 | 0.87 | 0.64*** | 0.90 | 0.70*** |
| | (0.21) | (0.14) | (0.09) | (0.12) | (0.09*) |
| Europe | 0.75*** | 0.66*** | 0.54*** | 0.70*** | 0.57*** |
| | (0.05) | (0.04) | (0.02) | (0.04) | (0.03) |
| North America | 1.58*** | 1.24 | 1.22 | 1.39* | 1.25 |
| | (0.22) | (0.18) | (0.21) | (0.26) | (0.24) |
| Oceania | 0.60** | 0.44*** | 0.40*** | 0.56** | 0.45*** |
| | (0.11) | (0.08) | (0.08) | (0.12) | (0.10) |
| South America | 0.79 | 0.61** | 0.51*** | 0.67** | 0.52*** |
| | (0.22) | (0.13) | (0.09) | (0.11) | (0.07) |
| Constant | 260013.70*** | 179112.10*** | 179774.70*** | 278514.10*** | 207195.00*** |
| | (14381.61) | (9734.03) | (5494.21) | (13315.21) | (8446.85) |
| Cragg & Uhler's $R^2$ | 0.588 | 0.653 | 0.666 | 0.666 | 0.687 |

*Note.* Robust standard errors are clustered by continent and reported in parentheses; *** p<0.01, ** p<0.05, * p<0.1.

**Table S2**

*Results for Negative Binomial regression. Dependent variable: International views (N = 1,220)*

| | M0: Baseline model | M1: Productivity | M2: International collab. | M3: P top 10% | M4: Full model |
|---|---|---|---|---|---|
| | IRR (SE) | IRR (SE) | IRR (SE) | IRR (SE) | IRR (SE) |
| P top 10% | | | | 1.86*** | 1.41*** |
| | | | | (0.25) | (0.06) |
| *P int collab.* | | | | | |
| Medium int. collab. | | | 1.86*** | | 1.71*** |
| | | | (0.18) | | (0.16) |
| High int. collab. | | | 4.87*** | | 2.67*** |
| | | | (0.84) | | (0.51) |
| *P* | | | | | |
| Medium productivity | | 1.58*** | | | 1.03 |
| | | (0.07) | | | (0.04) |
| High productivity | | 3.86*** | | | 1.08 |
| | | (0.44) | | | (0.12) |
| Pa f mf | 0.82* | 0.94 | 0.94 | 0.93 | 0.96 |
| | (0.08) | (0.01) | (0.07) | (0.08) | (0.06) |
| Dimensions WoS | 1.26 | 1.52*** | 1.56*** | 1.51*** | 1.61*** |
| | (0.19) | (0.17) | (0.16) | (0.15) | (0.14) |
| Age | 1.46*** | 1.29*** | 1.29*** | 1.27*** | 1.24*** |
| | (0.14) | (0.10) | (0.09) | (0.09) | (0.08) |
| Specialization bio | 1.03 | 0.80 | 0.85 | 0.68 | 0.71 |
| | (0.37) | (0.26) | (0.24) | (0.17) | (0.16) |
| Specialization eng | 0.98 | 1.13*** | 1.19*** | 1.08*** | 1.15*** |
| | (0.06) | (0.03) | (0.05) | (0.01) | (0.02) |
| Anglo-saxon country | 5.79*** | 6.13*** | 5.40*** | 4.60*** | 4.64*** |
| | (0.18) | (0.21) | (0.23) | (0.17) | (0.20) |
| *Continent* | | | | | |
| Africa | 1.6* | 1.38* | 0.89 | 1.40* | 0.99 |
| | (0.41) | (0.25) | (0.16) | (0.25) | (0.17) |
| Europe | 0.95 | 0.74*** | 0.55*** | 0.78** | 0.59*** |
| | (0.10) | (0.07) | (0.05) | (0.08) | (0.06) |
| North America | 1.74*** | 1.20*** | 1.18*** | 1.33*** | 1.19*** |
| | (0.12) | (0.06) | (0.04) | (0.12) | (0.06) |
| Oceania | 0.70*** | 0.44*** | 0.38*** | 0.56*** | 0.42*** |
| | (0.03) | (0.03) | (0.03) | (0.05) | (0.04) |
| South America | 0.40** | 0.24*** | 0.18*** | 0.27*** | 0.18*** |
| | (0.16) | (0.08) | (0.06) | (0.08) | (0.05) |
| Constant | 177761.90*** | 101992.5*** | 101887.30*** | 189460.60 | 121471.10*** |
| | (11944.49) | (6230.95) | (5885.00) | (14338.03) | (5680.53) |
| Cragg & Uhler's $R^2$ | 0.611 | 0.713 | 0.735 | 0.724 | 0.757 |

*Note.* Robust standard errors are clustered by continent and reported in parentheses; ***$p<0.01$, **$p<0.05$, *$p<0.1$.

**Table S3**

*Results for Negative Binomial regression. Curvilinear effects. Dependent variable: Total views (N = 1,220)*

|  | M1: Productivity | M2: International collab. | M3: Ptop 10% |
|---|---|---|---|
|  | IRR (SE) | IRR (SE) | IRR (SE) |
| P top 10% |  |  | 2.05*** |
|  |  |  | (0.13) |
| P top 10% srq |  |  | 0.96*** |
|  |  |  | (0.01) |
| P int collab. |  | 2.21*** |  |
|  |  | (0.19) |  |
| P int collab.*P int collab. |  | 0.94*** |  |
|  |  | (0.01) |  |
| P | 2.05*** |  |  |
|  | (0.11) |  |  |
| P*P | 0.95*** |  |  |
|  | (0.01) |  |  |
| Pa f mf | 0.92* | 0.93*** | 0.92** |
|  | (0.04) | (0.02) | (0.03) |
| Dimensions WoS | 1.53*** | 1.53*** | 1.53*** |
|  | (0.11) | (0.08) | (0.10) |
| Age | 1.25*** | 1.23*** | 1.27*** |
|  | (0.07) | (0.07) | (0.07) |
| Specialization bio | 0.82** | 0.86* | 0.78*** |
|  | (0.08) | (0.07) | (0.06) |
| Specialization eng | 0.92 | 0.95 | 0.88 |
|  | (0.09) | (0.10) | (0.13) |
| Anglo-saxon country | 2.31*** | 2.24*** | 1.90*** |
|  | (0.23) | (0.21) | (0.26) |
| Continent |  |  |  |
| Africa | 0.81 | 0.60*** | 0.75* |
|  | (0.13) | (0.09) | (0.11) |
| Europe | 0.93 | 0.69*** | 0.86* |
|  | (0.07) | (0.07) | (0.07) |
| North America | 1.38*** | 1.29*** | 1.40*** |
|  | (0.12) | (0.11) | (0.17) |
| Oceania | 0.57*** | 0.43*** | 0.62*** |
|  | (0.07) | (0.05) | (0.09) |
| South America | 0.56*** | 0.46*** | 0.57*** |
|  | (0.11) | (0.08) | (0.11) |
| Constant | 502401.50*** | 572904.40*** | 539023.00*** |
|  | (30128.72) | (44308.09) | (34936.98) |
| Cragg & Uhler's $R^2$ | 0.694 | 0.713 | 0.681 |

*Note. Results for Negative Binomial regression. Curvilinear effects. Dependent variable: Total views (N = 1,220)*